\documentstyle[12pt]{article}
\begin{document}
\baselineskip=24pt
\begin{center}
{\Large \bf General Relativity + Quantum Mechanics \\
$\Longrightarrow $ Discretized Momentum}\\
{\bf H. Gopalkrishna Gadiyar}\\
E-mail: padma@imsc.ernet.in
\end{center}

\vspace{2cm}
\begin{center}
{\bf Abstract}
\end{center}

The analogy between General Relativity and monopole physics is pointed
out and the presence of a 3-cocycle which corresponds to a source leads
to discretization of field momentum. This is analogous to the same
phenomena in monopole physics.

\newpage

Recently many attempts have been made to discretize spacetime, cutoff
momentum and hence improve quantum field theory. In this telegraphic
note we wish to draw an analogy between General Relativity and monopole
physics which leads to the same effect. 

Recall [1] that in monopole physics the triple commutator of velocity \\
$v~=~p-e A$, where $B= \nabla \times A$ is given by
$$
\left [ v^1, \left [v^2,v^3 \right ] \right ] + 
\left [ v^2, \left [v^3,v^1 \right ] \right ] + 
\left [ v^3, \left [v^1,v^2 \right ] \right ] = e \nabla \cdot B \, .
$$
 
In the presence of a monopole of strength $g$ located at the origin
$$
e \nabla \cdot B = 4 \pi g e \delta (r) \, .
$$
The failure of the Jacobi identity signals the occurance of a 3-cocycle
which leads to the quantization of $eg$. 

This can also be viewed in the language of differential forms as the
failure of the Bianchi identity for the curvature $F$. In the presence
of the monopole it is easily seen that the quantization at $eg$ follows.

The analogy with General Relativity [2] is based on the simple observation
that the equation relating curvature and the energy momentum tensor of
matter is 
$$
^{\ast} G = ^{\ast } T \, .
$$
As $d^{\ast}G = 0$ is a consequence of the contracted Bianchi identity
it follows that 
$$
d^{\ast}T=0
$$
which is the conservation of momentum. 

Let us for a moment assume that $d^{\ast}T \neq 0$. This would mathematically
signal the appearance of the 3-cocycle in analogy with the monopole case. 
In combination with the laws of quantum physics this would lead to
discretization of the corresponding `charge' which is 
$P^{\mu}= \int T^{\mu \nu} d\sigma ^{\nu}$. Hence the field momentum
takes discretized values. Hence in analogy with Dirac if there is a
source for energy momentum that would lead to discretization of field
momentum just as the presence of a single monopole would lead to charge
quantization. 

The author wishes to thank Professor H.S. Sharatchandra and Professor G.
Rajasekaran for discussions. He is grateful to Professor N.D. Haridass
who taught him modern differential geometry.

\begin{center}
{\bf References}
\end{center}

\begin{description}
\item [1.] There are many references to monopoles and cocycles. See for
example,\\
R. Jackiw, {\it Chern-Simons terms and cocycles in physics and mathematics},
Quantum field theory and quantum statistics, Essays in honor of the sixtieth
birthday of E.S. Fradkin, Editors: I.A. Batalin, C.J. Isham and G.A. 
Vilkovisky, Adam Hilger, 349-378.

\item [2.] We follow the standard notation. See for example, \\
Charles W. Misner, Kip S. Thorne and John A. Wheeler,
{\it Gravitation}, W.H. Freeman \& Company, 1970.
\end{description}
\end{document}